\documentclass[twocolumn,preprintnumbers,amsmath,amssymb,nobibnotes,nofootinbib,superscriptaddress]{revtex4}
\usepackage[utf8]{inputenc}
\usepackage{graphicx}
\usepackage{dcolumn}
\usepackage[usenames,dvipsnames]{xcolor}
\RequirePackage[colorlinks=true
,urlcolor=blue
,anchorcolor=blue
,citecolor=blue
,filecolor=blue
,linkcolor=blue
,menucolor=blue
,pagecolor=blue
,linktocpage=true 
,pdfproducer=medialab
]{hyperref}
\usepackage{amsmath}
\usepackage{mathrsfs}
\usepackage{slashed}
\usepackage{cancel}

\usepackage{feynmp}
\makeatletter
\def\endfmffile{%
  \fmfcmd{\p@rcent\space the end.^^J%
          end.^^J%
          endinput;}%
  \if@fmfio
    \immediate\closeout\@outfmf
  \fi
  \ifnum\pdfshellescape=\@ne
    \immediate\write18{mpost \thefmffile}%
  \fi}
\makeatother

\newcommand{\LL}{\mathscr{L}}
\newcommand{\cG}{\mathscr{G}}
\def\cB{{\cal B}}
\def\cC{{\cal C}}

\def\cG{{\cal G}}

\def\cO{{\cal O}}

\def\cT{{\cal T}}

\def\Tr{{\rm Tr}}

\def\be{\begin{equation}}
\def\ee{\end{equation}}
\def\beq{\begin{equation}}
\def\eeq{\end{equation}}
\def\bc{\begin{center}}
\def\ec{\end{center}}
\def\bea{\begin{eqnarray}}
\def\eea{\end{eqnarray}}

\def\bry{\begin{array}}
\def\ery{\end{array}}

\def\nn{\nonumber}
\def\nt{\noindent}

\newcommand{\hc}{\text{h.c.}}
\def\fA{\mbox{{\bf M$_{4+5}$}}}
\def\fB{\mbox{{\bf M$_{4+14}$}}}
\def\oA{\mbox{{\bf M$_{1+5}$}}}
\def\oB{\mbox{{\bf M$_{1+14}$}}}

\def\fourpletL{\Psi_{\textbf{4}L}}
\def\fourpletR{\Psi_{\textbf{4}R}}
\def\fourplet{\Psi_{\bf{4}}}

\def\singletL{\Psi_{\textbf{1}L}}
\def\singletR{\Psi_{\textbf{1}R}}
\def\singlet{\Psi_{\bf{1}}}

\def\5qplet{q_L^{\mathbf{5}}}
\def\qBar5plet{\overline{q}^{\bf{5}}_L}
\def\14qplet{q_L^{\bf{14}}}
\def\q14Barplet{\overline{q}^{\bf{14}}_L}

\def\u5plet{u_R^{\bf 5}}
\def\uBar5plet{\overline{u}^{\bf{5}}_R}
\def\usinglet{u_R^{\bf 1}}

\def\d5plet{d_R^{\bf 5}}
\def\dBar5plet{\overline{d}^{\bf{5}}_R}

\def\Xtt{{X_{\hspace{-0.09em}\mbox{\scriptsize2}\hspace{-0.06em}{\raisebox{0.1em}{\tiny\slash}}\hspace{-0.06em}\mbox{\scriptsize3}}}}
\def\Xft{{X_{\hspace{-0.09em}\mbox{\scriptsize5}\hspace{-0.06em}{\raisebox{0.1em}{\tiny\slash}}\hspace{-0.06em}\mbox{\scriptsize3}}}}

\newcommand{\ie}{i.e.\,\,}

\newcommand{\hhref}[1]{\href{http://arxiv.org/abs/#1}{arXiv:#1}}

\begin{document}
%
%

\title{Top partners tackling vector dark matter}

\author{Juan Yepes}
\email{juan.yepes@usm.cl}
\affiliation{
Department of Physics and Centro Cient\'{i}fico-Tecnol\'{o}gico de Valpara\'{i}so\\
Universidad T\'{e}cnica Federico Santa Mar\'{i}a, Valpara\'{i}so, Chile
}

\begin{abstract}
\nt The WIMP-nucleon scattering cross section in a simple dark matter model and its constraints from the latest direct detection experiment are treated here at the loop level. We consider a scenario with an emerging vector dark matter field that interacts with the Standard Model quarks, via loop contributions that are sourced from a scalar mediator. The involved parameter space for the dark matter-mediator masses is constrained by the Xenon1T limit and the neutrino floor. The current direct detection bounds are eluded by invoking the top partners in a Composite Higgs model, whose scale mass helps us in suppressing the WIMP-nucleon cross section.
\end{abstract}
\maketitle

\section{Introduction}
\nt The dark matter (DM) stands as one of the most tantalising components in our Universe, becoming nowadays one of the most mysterious conundrums unknown so far. The particle physics and astrophysics community are currently seeking for DM signals in order to unveil its nature and involved properties. A weak interacting massive particle (WIMP)~\cite{Steigman:1984ac} has been a long-standing DM candidate, having resurged since the successful experimental confirmation of the Standard Model (SM) Higgs sector. Singlet elementary scalars have been additionally considered in the last years~\cite{Djouadi:2011aa,Mambrini:2011ri,Kadastik:2011aa,
Batell:2011pz,He:2011gc}, extended a posteriori with composite scalars in the light of composite Higgs models (CHMs). These scenarios represent an appealing option to supersymmetry, rendering the DM candidates much lighter than the new physics compositeness scale, as they enter onto the stage like the pseudo-Nambu-Goldstone bosons (pNGBs) of a new strongly interacting sector. Specific models have been explored for instance in~\cite{Frigerio:2012uc, Gripaios:2009pe,Chala:2012af,Barnard:2014tla}. In other contexts, there have been interesting claims for vector dark matter (VDM), arising from an additional $U(1)_X$ gauge symmetry~\cite{Hambye:2008bq,Lebedev:2011iq,Farzan:2012hh,Baek:2012se,
Baek:2014jga,Duch:2015jta}, recently analysed and compared against scalar dark matter (SDM) models~\cite{Azevedo:2018oxv}.

On the other hand, different WIMP-ordinary matter interactions exist, splitting thus the WIMP searches into three methods: direct detection (DD), indirect detection (ID) and collider search. Among them, the DD experiments pursue the scattering of dark matter particles off atomic nuclei. No clear signals of dark matter from DD experiments have been detected till now, leaving us instead with upper limits on the WIMP-nuclei scattering cross section. The coherent sum of spin-independent (SI) nucleon amplitudes leads to a strong enhancement for large nuclei. Enhanced by the squared total nucleon number in a nucleus, the loop level SI amplitudes can compensate the loop suppression, dominating thus the WIMP-nuclei cross section. No such enhancement for SD interactions occurs in nuclei. Hence, recent upper limits from DD experiments should sizeably constrain the loop-induced SI cross section.

We assume in this work a framework consisting of complex or real vector DM candidates $X_\mu$, interacting with the SM particles via the scalar mediator $\eta$. We compute the loop level SI cross section for this simplified VDM model, exploring as well the constraints imposed by the latest direct detection experiments~\cite{Tan:2016zwf,Aprile:2017iyp}. The DD bound will be compared with those from ID and collider search in a future work. This approach has been thoroughly used for analysing dark matter searches in DD, ID and collider experiments~\cite{Buchmueller:2013dya,Arina:2014yna,Alves:2015pea,
Abdallah:2015ter,Boveia:2016mrp,Arina:2016cqj,Ismail:2016tod,
Li:2016uph,Balazs:2017hxh,Li:2017nac,Morgante:2018tiq}.

\section{Effective Lagrangian}
\label{EffectiveLagrangian}

\nt In this simple scenario, the dark matter annihilation occurs through the s-channel exchange of either a spin-0 or spin-1 mediator. The latter case has been constrained by the searches of $Z'\to$ dijet at the Large Hadron Collider (LHC)~\cite{Khachatryan:2016ecr,Sirunyan:2016iap,Sirunyan:2017nvi,
Sirunyan:2018wcm}. The constraints from direct detection experiments rule out the scenarios in which the DM annihilates through a spin-0 mediator coupled to  SM scalar quark currents~\cite{Berlin:2014tja}. We assume henceforth the scalar mediator scenario\footnote{The spin-0 mediator case is generically suppressed by the Yukawa couplings via the minimal flavor violation prescription, the which suppresses the scalar bilinear $\bar{q}q$ and the pseudo-scalar one $\bar{q}\gamma_5 q$ by the SM quark mass $m_q$~\cite{Buras:2000dm,Goodman:2010ku}. In here the scalar mediator will couple to the SM fermions proportionally to their mass as it can be seen later on.}
\be
\LL_{\rm X} = -g_{X\eta} M_X \,X^\mu X_\mu\,\eta -  i\eta\sum_{q}g_{\eta q} \bar{q}\gamma_5 q.
\label{VDM-scalar}
\ee

\nt These interactions entail unsuppressed low-velocity annihilation cross section ($\sigma v \sim 1$), and therefore are able to account for the observed gamma-ray excess~\cite{Berlin:2014tja}.
On the other hand, the presence of a scalar coupling $g^S_{\eta q}\,\eta\,\bar{q}q$ in~\eqref{VDM-scalar}, would induce momentum-suppressed SI interactions already at tree-level, motivating thus the calculation of the non-suppressed SI WIMP-nucleon cross sections from loop diagrams. This motivates even further to focus on the pseudoscalar terms of~\eqref{VDM-scalar} and its loop contributions, although a complete theoretical scenario requires the presence of both scalar and pseudoscalar couplings. Including the former ones, leads to the tree level amplitude  $\frac{M_X}{M^2_\eta}g_{X\eta}g^S_{\eta q}$, and the required tuning to play a subdominant role would be $g^S_{\eta q} \ll \frac{M^2_\eta}{M_X}g_{X\eta}\,g^2_{\eta q}\,\mathcal{C}^q_{\text{loop}}$, with $\mathcal{C}^q_{\text{loop}}$ encoding the implied loop contribution triggered by the pseudoscalar terms. Hereinafter we will only assume the terms in~\eqref{VDM-scalar}, leaving the scalar-pseudoscalar scenario for a future analysis.

The analysis of dark matter searches in the literature commonly assume $g_{X\eta}=g_{\eta q}=1$. Under this assumption, these dark matter models are described by two parameters: the dark matter and the mediator masses $M_X$ and $M_\eta$ respectively. Nonetheless, such coupling $g_{\eta q}$ might appear in different frameworks, as in the case of CHM scenarios. In this work it is invoked the minimal global symmetry $\cG=SO(5)$~\cite{Agashe:2004rs}, spontaneously broken to $SO(4)$ by the strong sector at the scale $f$. Furthermore, the elementary sector fields, together with the top partners $\Psi$ that transform in the unbroken $SO(4)$, are implemented to write the following mass mixing terms
\be
\begin{aligned}
\LL_{\text{mix}} =& y_L f \left(\qBar5plet\,U\right)_i \left(\fourpletR\right)^{i} + y_R f \left(\uBar5plet\,U\right)_i \left(\fourpletL\right)^{i} + \hc +,\\[5mm]
& + \tilde{y}_L f \left(\qBar5plet\,U\right)_5\singletR + \tilde{y}_R f \left(\uBar5plet\,U\right)_5\singletL + \hc
\label{fA-oA-mix}
\end{aligned}
\ee

\nt and 
\be
\begin{aligned}
\LL'_{\text{mix}}=& y_L f\left(U^t\q14Barplet U\right)_{i\,5} \left(\fourpletR\right)^{i} + \tilde{y}_L f\left(U^t\q14Barplet U\right)_{5\,5} \singletR\,+ \\[5mm]
& +y_R\,f\left(U^t\,\q14Barplet\,U\right)_{5\,5} \,\usinglet + \hc
\label{fB-oB-mix}
\end{aligned}
\ee

\nt with the pNGB fields encoded in the $5\times 5$ Goldstone matrix $U$, defined as
\be
U=\exp\left[i \frac{\sqrt{2}}{f}\,\Pi^i\,T^i\right],
\label{GB-matrix}
\ee

\nt with $T^i$ the coset $SO(5)/SO(4)$-generators (App.~\ref{CCWZ}), where $\Pi^i$ are the pNGB fields and $f$ the decay constant. Small mixing couplings $y_{L(R)}$ and $\tilde{y}_{L(R)}$ trigger the Goldstone Boson (GB) symmetry breaking. The top partners fourplet $\fourplet$ and singlet $\singlet$ are embedded in the unbroken $SO(4)$ as
\be
\fourplet={1\over \sqrt{2}}\left(\begin{matrix}
i\cB-i\Xft\\
\cB+\Xft\\
i\cT+i\Xtt\\
-\cT+\Xtt
\end{matrix}\right),\qquad\qquad \singlet=\widetilde{\cT}\,.
\label{fourplet-singlet}
\ee

\nt One SM-like quark doublet $(\cT,\cB)$, plus an exotic $7/6$-hypercharge $(\Xft,\Xtt)$ are contained in $\fourplet$, while $\singlet$ contains only one exotic top-like state $\widetilde{\cT}$. The elementary sector is prescribed by the partial compositeness mechanism through the GB symmetry breaking terms  $y\,\bar q\,\cO_q$, with the strong sector operator $\cO_q$ transforming in one of the $SO(5)$-representations. We assume here two elementary sector embeddings: either as a fundamental ${\bf 5}$
\be
\5qplet={1\over \sqrt{2}}\left(
i d_L,\,\,
 d_L,\,\,
i u_L,\,\,
- u_L,\,\,
0
\right)^T,
\label{left-5}
\ee
\be
\u5plet = \left(
0,\,\,
0,\,\,
0,\,\,
0,\,\,
u_R
\right)^T,
\label{right-5}
\ee

\nt or the ${\bf 14}$ representation
\be
\hspace*{0.5cm}
\14qplet={1\over \sqrt{2}}\left(\begin{matrix}
0 & 0 & 0 & 0 & i d_L\\
0 & 0 & 0 & 0 & d_L\\
0 & 0 & 0 & 0 & i u_L\\
0 & 0 & 0 & 0 & -u_L\\
i d_L & d_L & i u_L & -u_L &0\\
\end{matrix}\right),
\quad
\usinglet\,.
\label{left-right-14}
\ee

\nt In the former scenario, both fermion chiralities have elementary representatives coupled to the strong sector through ${\bf 5}$-plets, whereas in the latter the right-handed quark enters as a totally composite state. All these matter shape four models, here denoted as $\bf M_{\Psi +q}=\{\fA,\,\fB,\,\oA,\,\oB\}$, with the first two hereinafter referred as the fourplet models, while the last two as singlet scenarios.

The aforementioned fermionic matter is coupled to the scalar resonance $\eta$, here described as a singlet of $SU(2)_L\times SU(2)_R$, and previously considered in CHMs~\cite{Contino:2011np}. In fact, in the ${\bf 5}$-plets scenarios coupled to $\eta$, the pseudoscalar couplings can be retrieved from~\cite{Norero:2018dfg}
\be
\begin{aligned}
\LL_{q\psi\eta} &= \eta\,\Bigl[y_{q\psi}\left(\qBar5plet U\right)_i\left(\fourpletR\right)^{i} + y_{u\psi}\left(\uBar5plet U\right)_i\left(\fourpletL\right)^{i} + \\[5mm]
& + \tilde{y}_{q\psi} \left(\qBar5plet U\right)_5\singletR + \tilde{y}_{u\psi}\left(\uBar5plet U\right)_5\singletL\Bigr]+\hc,
\label{fA-oA-mix-eta}
\end{aligned}
\ee

\nt with $y_{q(u)\psi}$ and $\tilde{y}_{q(u)\psi}$ controlling the scalar-fermion interactions. These couplings are responsible for the fermionic decays of $\eta$ into the pure SM final states, giving rise also to decay channels with a single or double partner field. For the ${\bf 14}$-plets models we have similar Lagrangians 
\be
\begin{aligned}
&\LL'_{q\psi\eta}=\\[3mm]
&\eta \Bigl[y_{q\psi}\left(U^t\q14Barplet U\right)_{i\,5} \left(\fourpletR\right)^{i}  + y_{qu}\left(U^t\q14Barplet U\right)_{5\,5} \,\usinglet\,\,+   \\[5mm]
& + \tilde{y}_{q\psi}\left(U^t\q14Barplet U\right)_{5\,5}\singletR\Bigr] \,\,+\,\,\hc
\label{fB-oB-mix-eta}
\end{aligned}
\ee

\nt Via the equations of motion of the top partners fields  (App.~\ref{Partners-EOM}), the Lagrangians in~\eqref{fA-oA-mix-eta} and~\eqref{fB-oB-mix-eta} lead to a set of interactions, the which couple the SM top quarks to the spin-0 mediator. Their associated couplings are appropriately mapped onto each model as
\beq
\small{
\begin{aligned}
&\fA:\quad g_{\eta q} =\sqrt{{\xi\over 2}}\,\frac{ \left(\eta _R\operatorname{Im}(y_{q \psi }) - \eta_L\left(\eta _L^2+1\right)\operatorname{Im}(y_{u \psi })\right)}{\left(\eta _L^2+1\right){}^{3/2}},\\
&\fB:\quad g_{\eta q} =\sqrt{2\,\xi}\,\frac{ \left(i\,\eta _L \eta _R\operatorname{Re}(y_{q \psi })-\left(\eta _L^2+1\right)\operatorname{Im}(y_{q u})\right)}{\left(\eta _L^2+1\right){}^{3/2}},\\
&\oA:\quad g_{\eta q} =\sqrt{{\xi\over 2}}\,\frac{ \left(\tilde{\eta }_L\operatorname{Im}(\tilde{y}_{u \psi })-\tilde{\eta }_R\left(\tilde{\eta }_R^2+1\right)\operatorname{Im}(\tilde{y}_{q \psi })\right)}{\left(\tilde{\eta }_R^2+1\right)^{3/2}},\\
&\oB:\quad g_{\eta q} =-\sqrt{2\,\xi}\sqrt{\tilde{\eta }_R^2+1}\,\operatorname{Im}(y_{q u})
\end{aligned}
}
\label{Couplings-fRfReta}
\eeq

\nt where $\xi=v^2/f^2$ with $v=246$ GeV, and the coefficients $\eta _{L(R)} \equiv  y_{L(R)} \mathit{f}/M_{\bf{4}}$ and $\tilde{\eta}_{L(R)}\equiv \tilde{y}_{L(R)} \mathit{f}/M_{\bf{1}}$ (App.~\ref{Partners-EOM}). The mass scales $M_{\bf{4(1)}}$ are associated to the fermionic resonances $\Psi_{\bf{4(1)}}$. They are commonly parametrised through the generic coupling $g_\Psi$ as $M_{\bf{4}}=M_{\bf{1}}=M_\Psi \equiv g_\Psi  f$. Notice how the top partner mass dependence is directly exhibited through the parameters $\eta_{L(R)}$ and $\tilde{\eta}_{L(R)}$, while it is also manifested via the parameter $\xi$ and the relation $M_\Psi =g_\Psi  f$. When exploring the cross sections and their sensitivity to the variations of the top partner mass, two options appear: 
\begin{itemize}

\item either to vary the top partner mass scale while properly changing the coupling $g_\Psi$, and conserving thus $\xi$ at a fixed value, or

\item just to vary $M_\Psi$ while keeping fixed $g_\Psi$, implying therefore a varying $\xi$.

\end{itemize}

\nt The second option effectively suppresses much more the couplings in~\eqref{Couplings-fRfReta}, rather than the first one. This feature will be reflected later on in the computation of the WIMP-nucleon cross section. 

\section{Cross sections and constraints}
\label{Cross-sections-constraints}

\begin{figure}
\begin{center}
\begin{tabular}{cc}
\input{Fdiagrams/WIMP-scattering-Box-Diagram}&
\hspace*{0.5cm}
\input{Fdiagrams/WIMP-scattering-Box-Diagram-extra}
\end{tabular}
\caption{Box diagrams generating the SI WIMP-nucleon cross section. The diagrams arise from the interactions in~\eqref{VDM-scalar}.}
\label{Box-diagrams}
\end{center}
\end{figure}

\begin{figure}
\begin{center}
\begin{tabular}{cc}
\input{Fdiagrams/WIMP-scattering-Triangle-Diagram}&
\hspace*{0.5cm}
\input{Fdiagrams/WIMP-scattering-Two-loop-Diagram}
\end{tabular}
\caption{Triangle diagram contributing to the DM-quark cross section and the two-loop triggering the DM-gluon interaction.}
\label{Triangle-Two-Loop-diagrams}
\end{center}
\end{figure}

\nt The Lagrangian in~\eqref{VDM-scalar} leads to an effective low energy scenario that entails WIMP-quark scattering amplitudes via the one and two-loop box diagrams as it is shown in Fig.~\ref{Box-diagrams}-\ref{Triangle-Two-Loop-diagrams}, and schematically written for this model as~\cite{Glaus:2020ape}
\be
\begin{aligned}
&\LL^{\text{eff}} =\\
&\sum_{q=u,d,s}\left(f_q\,X^\mu X_\mu\,m_q\,
          \bar q\,q  \,+\,
          \frac{g_q}{M^2_X}\,X^\rho i \partial^{\mu} i \partial^{\nu}X_\rho\,
         \cO^q_{\mu\nu}\right)\,+\, \\[1.5mm]
&\,+\, f_G\,X^\rho X_\rho\,G^{a}_{\mu\nu}\,G^{a \,\mu\nu}\
\label{effective-low}
\end{aligned}
\ee

\nt with $G_{\mu\nu}^a$ ($a=1,...,8$)  the gluon field strength tensor and $\mathcal{O}_{\mu\nu}^q$  the quark twist-2 operator~\cite{Hisano:2010ct,Hisano:2015bma}
\be
    \mathcal{O}^q_{\mu\nu} = \bar{q}\left(\frac{i \partial_\mu \gamma_\nu + i \partial_\nu \gamma_\mu}{2} - \frac{1}{4} g_{\mu\nu} i \slashed{\partial}\right)q\,.  
\ee

\nt As soon as an effective mediator-Higgs coupling is brought into the model, a triangle diagram appears like the one in Fig.~\ref{Triangle-Two-Loop-diagrams} that contributes to the Wilson coefficient $f_q$. Such coupling will be justified later on.  

The box diagram, on the other hand, triggers contributions for $f_q$ and $g_q$. Accounting for the two-loop diagram in Fig.~\ref{Triangle-Two-Loop-diagrams}, a non-zero contribution arises for $f_G$. After transforming the WIMP-quark amplitude onto the WIMP-nucleon amplitude, the SI DM-nucleon cross section turns out to be
\be
\sigma_{{\rm DM} N}^{\rm SI}={1\over \pi}\left(\frac{m_N}{M_X+m_N}\right)^2|\cC_N|^2
\label{Amplitude-cross-section}
\ee

\nt where $m_N$ is the nucleon mass and
\be
\frac{\cC_N  }{m_N} = \sum_{q=u,d,s} f'_q f^N_q \,+\, \sum_{q\neq t} \frac{3}{4}\left(q^N(2)+\bar q^N(2)\right)g'_q +\frac{2}{27}
    f^N_G f'_G
\label{final-Amplitude}
\ee

\nt with 
\be
f'_q = f^{\text{tri}}_q \,+\, f^{\text{box}}_q,\quad
g'_q =g^{\text{box}}_q,\quad
f'_G =f^{\text{tri}}_G\,+\, f^{\text{box}}_G
\label{final-Wilson}
\ee

\nt and the superscript indicating the type of loop contribution for each coefficient. These are listed in App.~\ref{Loop-function}. The nuclear form factors $f^N_{q,G}$ in~\eqref{final-Amplitude} are taken from \texttt{micrOmegas}~\cite{Belanger:2018mqt}, while the second moments of the quark parton distribution functions $q^N(2)$ and $\bar{q}^N(2)$  from~\cite{Abe:2018emu,Pumplin:2002vw}. 

The energy scale relevant for DM direct detection experiments is lower than the charm, bottom and top quark masses. After integrating them out of the theory, a coefficient $f^{\text{tri}}_G$ emerges from the triangle heavy-quark loop that gives rise to the effective DM-gluon interactions via~\cite{Shifman:1978zn}
\be
\label{QQGGShifman}
m_Q \bar{Q} Q\quad\Rightarrow\quad -\frac{\alpha_s}{12\pi} G^{a}_{\mu\nu} G^{a\mu\nu}
\ee

\nt with $Q=c,b,t$. The box diagram sources $f^{\text{box}}_G$ through the two-loop diagram in Fig.~\ref{Triangle-Two-Loop-diagrams}. In this case, the procedure is less direct than the momentum expansion done for the box diagram contributing to $f^{\text{box}}_q$ and $g^{\text{box}}_q$. In fact, such expansion is no longer valid for $m_Q  \gg  M_\chi, M_\eta$ as it is also argued in~\cite{Abe:2018emu}. When $m_Q  \ll  M_\chi, M_\eta$ the two-loop computation cannot be simplified by first integrating out the mediator $\eta$ and then using~\eqref{QQGGShifman}. Otherwise the contribution from an additional two-loop diagram similar to that in Fig.~\ref{Triangle-Two-Loop-diagrams} would be missing (not shown here), and hence inducing  large errors a posteriori~\cite{Abe:2018emu}. Thus, it is mandatory to calculate the two-loop diagrams in Fig.~\ref{Triangle-Two-Loop-diagrams}, using the Fock-Schwinger gauge for the gluon field~\cite{Novikov:1983gd}, and extracting afterwards the effective DM-gluon operator. Such calculation is done in here following the procedure outlined in~\cite{Ertas:2019dew}. The result is reported in App.~\ref{Loop-function}.

All the couplings $g_{\eta q}$ for the down quark sector are vanishing as the associated masses remain zero in this scenario. This is due to the absence of the right handed down quark fields, see~\eqref{left-5}-\eqref{left-right-14}. Therefore, only the up quark-like couplings $g_{\eta q}$ are left in~\eqref{final-Wilson}. $f'_q$ and $g'_q$ are found to be one order of magnitude smaller than $f'_G$, if $g_{\eta q}=1$ is assumed in the DM mass range 20-10000 GeV and with a mediator mass of 100 GeV.  Now let us use the couplings of~\eqref{Couplings-fRfReta} by implementing the setting $\operatorname{Im}(y_{q \psi })=\operatorname{Im}(y_{u \psi })=\operatorname{Re}(y_{q \psi })=\operatorname{Im}(y_{q u})=m_q/v$ and $\operatorname{Im}(\tilde{y}_{u \psi })=\operatorname{Im}(\tilde{y}_{q \psi })=m_q/v$ (consistent with MFV), where $q=u,c,t$. It is found in this case that the first and second term in~\eqref{final-Wilson} are $\sim 10^{-4}$–$10^{-3}$ smaller than the third one in the same DM mass range. Including all the terms in the amplitude does not affect substantially the final SI cross section\footnote{According to~\cite{Dienes:2013xya}, the different weightings for the contributions from the light quarks were considered in the resulting dark-matter/nucleon couplings, and for their associated dark-matter scattering rates. As it was mentioned before, the final SI cross section is not substantially altered when including all the contributions from the light quarks.}.

On the other hand, when including the scalar coupling $g^S_{\eta q}\,\eta\,\bar{q}q$ in~\eqref{VDM-scalar}, tree-level contributions emerge for the SI interactions. In this case, the contribution from light quarks receives an enhancement proportional to $\frac{m_p}{m_q}$, while at the amplitude level one has  $\frac{M_X}{M^2_\eta}g_{X\eta}g^S_{\eta q}$. For the universal quark couplings, one has a tree level amplitude roughly of value 0.028 from the light quarks, while considering a non-universal situation one gets $1.25\times 10^{-6}$. This highlights the importance of the scalar quark coupling for the universal case, with contributions no longer negligible. Nonetheless, as it was stated before, the tuning $g^S_{\eta q} \ll \frac{M^2_\eta}{M_X}g_{X\eta}\,g^2_{\eta q}\,\mathcal{C}^q_{\text{loop}}$ is invoked in order to disregard tree-level momentum-suppressed SI interactions. 

Notice that in our work there is no  triangle diagram contributing to the SI DM-nucleon amplitude from the Lagrangian in~\eqref{VDM-scalar}. Nonetheless, additional diagrams at the loop level might emerge if gauge invariant $\eta$-$\eta$-$h$ interactions are included, with $h$ the physical Higgs field. Indeed, following~\cite{Arcadi:2017wqi,Sanderson:2018lmj}, it is possible to introduce a proper scalar potential mixing the mediator and the Higgs field. An effective coupling $\eta$-$\eta$-$h$ is obtained after diagonalising the scalar sector, leading thus to an additional triangle diagram that contributes to the SI cross section. In the case of~\cite{Arcadi:2017wqi,Sanderson:2018lmj}, such contribution entails additional parameters of the underlying scenario (the Higgs mass and the mass of extra scalar introduced by the set-up), apart from those previously implied by the model (the vector dark matter mass and the initial scalar mediator). As an example of a feasible realisation in our minimal framework, a Higgs potential coupled to the mediator is constructed in App.~\ref{Triangle-diagram}, where it has been sketched the contribution induced by the triangle diagram to the SI cross section. Such contribution is included throughout the next plots and constraints.

\begin{figure}
\begin{center}
\includegraphics[scale=0.355]{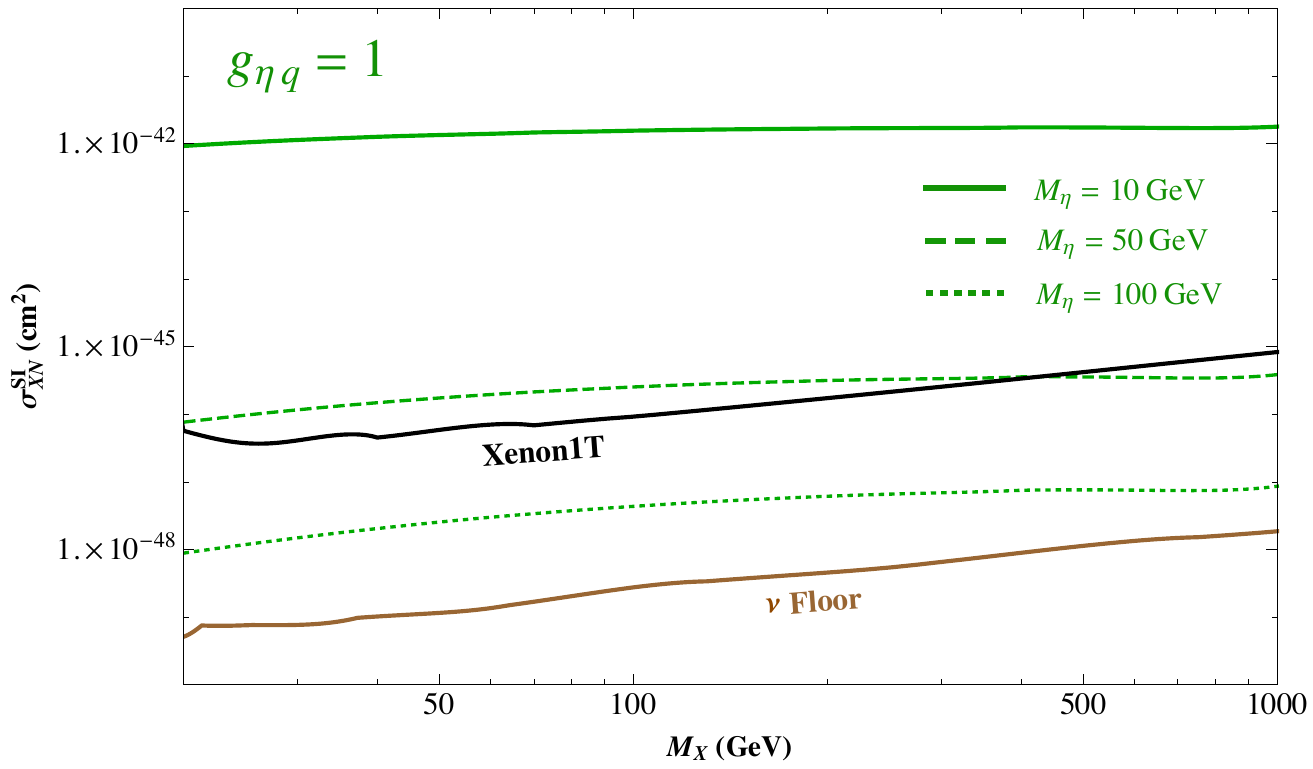}
\caption{\sf SI cross section at 1-loop versus vector DM mass, for the generic example $g_{\eta q}=1$ without quark mass proportionality assumed, and the mediator masses $M_\eta=10,\,50,\,100$ GeV (thick, dashed, dotted curves). The Xenon1T upper limit (thick black) can be partially/totally eluded, while the neutrino floor (brown) is still below the cross sections. See the text for details.}
\label{Cross-sections-generic-coupling}
\end{center}
\end{figure}

The different components of the loop coefficient $\cC_N$ in~\eqref{final-Amplitude}, are reported in App.~\ref{Loop-function}. The resulting Passarino-Veltman functions~\cite{Passarino:1978jh} are further reduced via Package-X~\cite{Patel:2015tea}. As a first approach, let us consider the generic example
with $g_{\eta q}=1$ in Fig.~\ref{Cross-sections-generic-coupling} (no quark mass proportionality assumed). We display the SI cross section versus the DM mass, assuming $g_{X\eta}=1$ and mediator masses $M_\eta=10,\,50,\,100$ GeV (thick, dashed, dotted curves). Notice how the heavier mediator reduces the WIMP-nucleon cross section, while the Xenon1T upper limit (thick black) is partially evaded for a mediator of 50 GeV at higher DM masses, being totally eluded in the whole DM mass region with a heavier mediator of 100 GeV. The cross sections are still above the neutrino floor (brown), the which could be evaded for a heavier mediator.

\begin{figure}
\begin{center}
\includegraphics[scale=0.36]{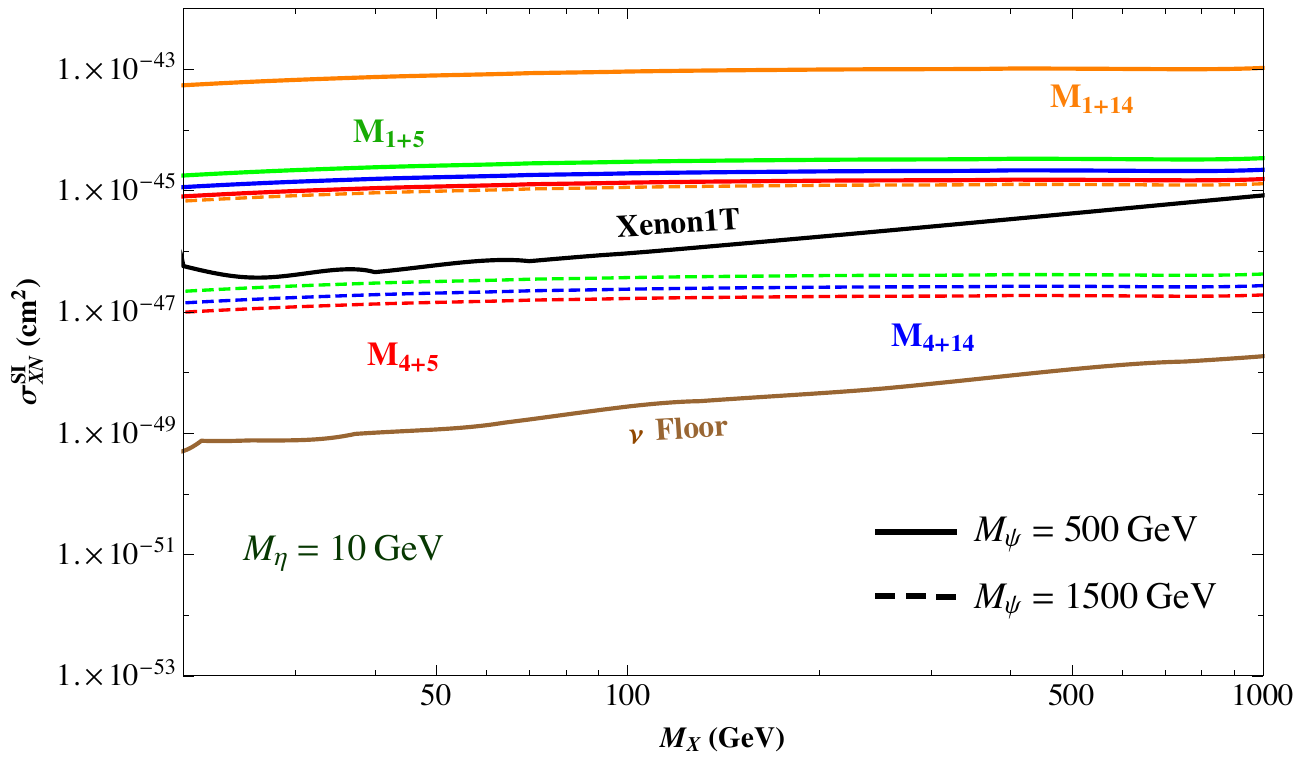}\\
\vspace*{0.5cm}
\includegraphics[scale=0.36]{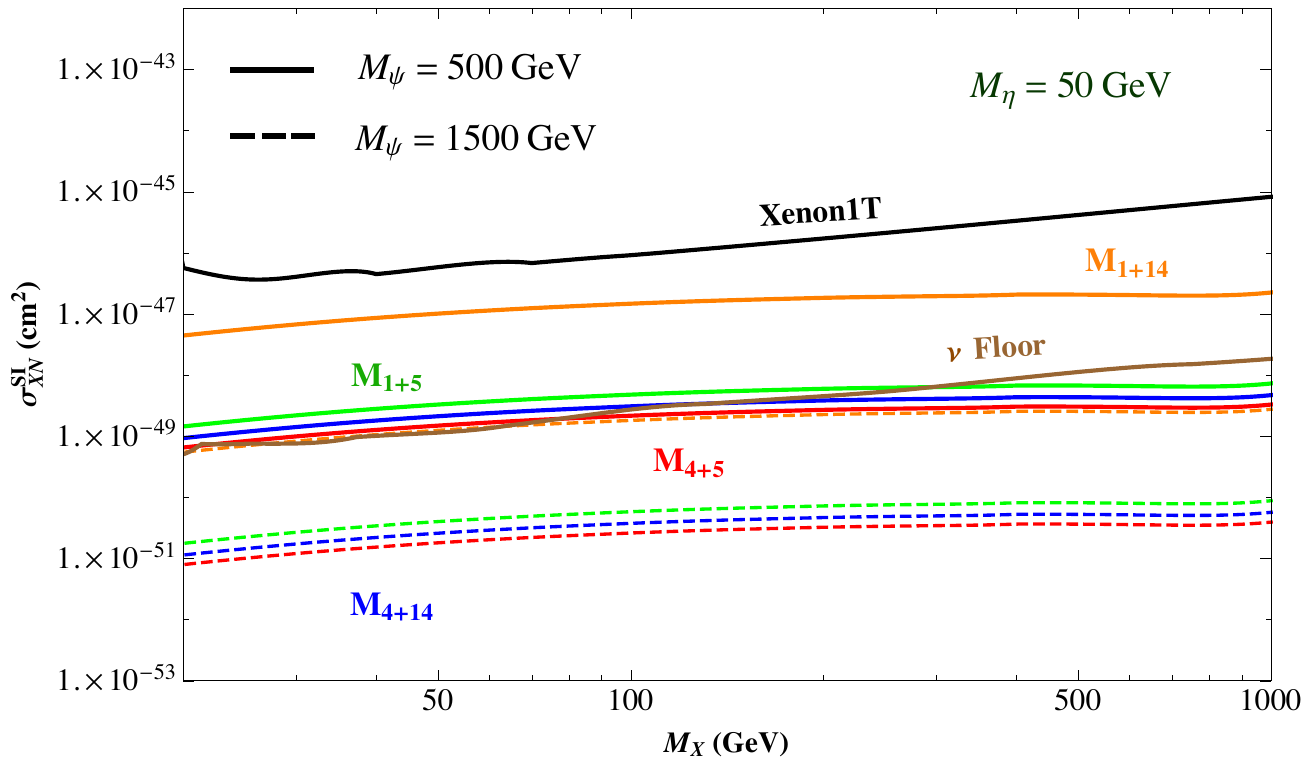}\\
\vspace*{0.4cm}
\includegraphics[scale=0.36]{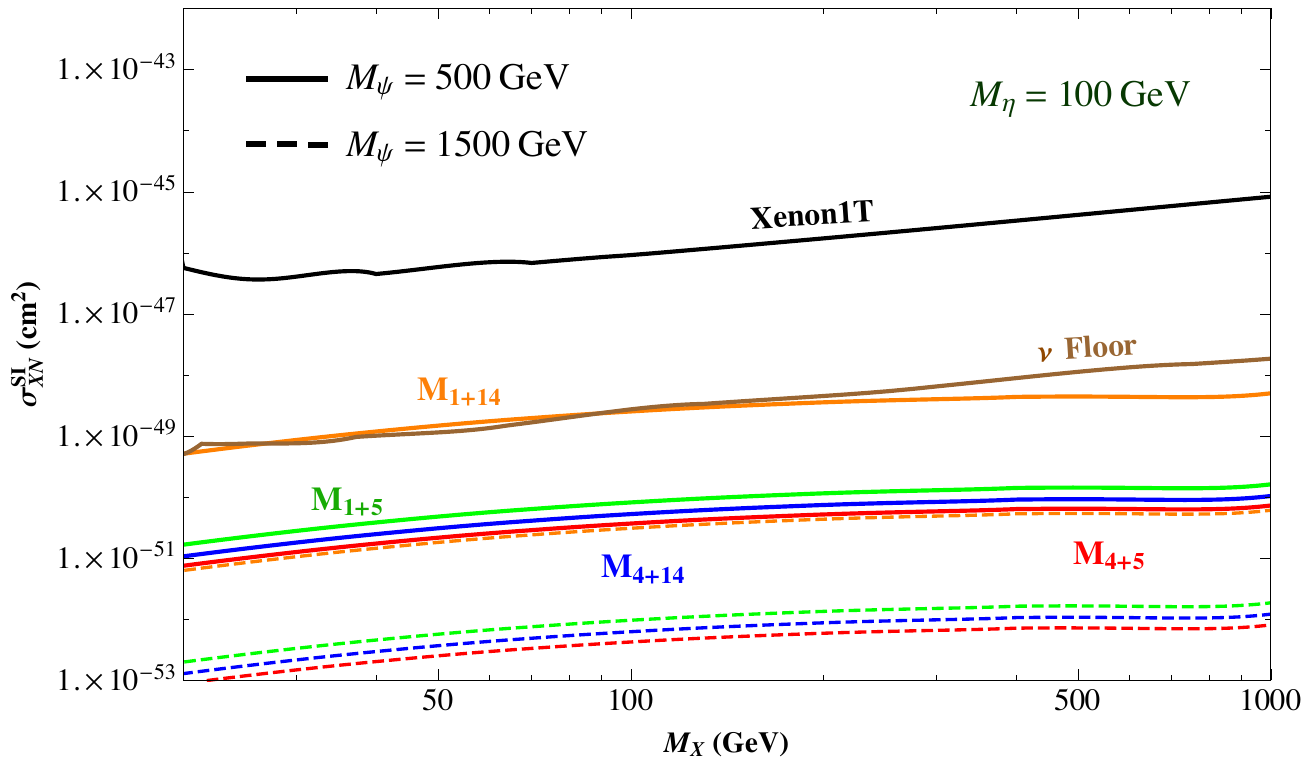}
\caption{\sf Implementing the couplings of~\eqref{Couplings-fRfReta}, the Xenon1T upper limit (thick black), and even the neutrino floor (brown), can be evaded in some of the models for a fixed mediator mass $M_\eta=10,\,50,\,100$ GeV (upper-centred-lower plots) and $M_\Psi=500,\,1500$ GeV (thick-dashed). See the text for details.}
\label{Cross-sections}
\end{center}
\end{figure}

\begin{figure}
\begin{center}
\includegraphics[scale=0.355]{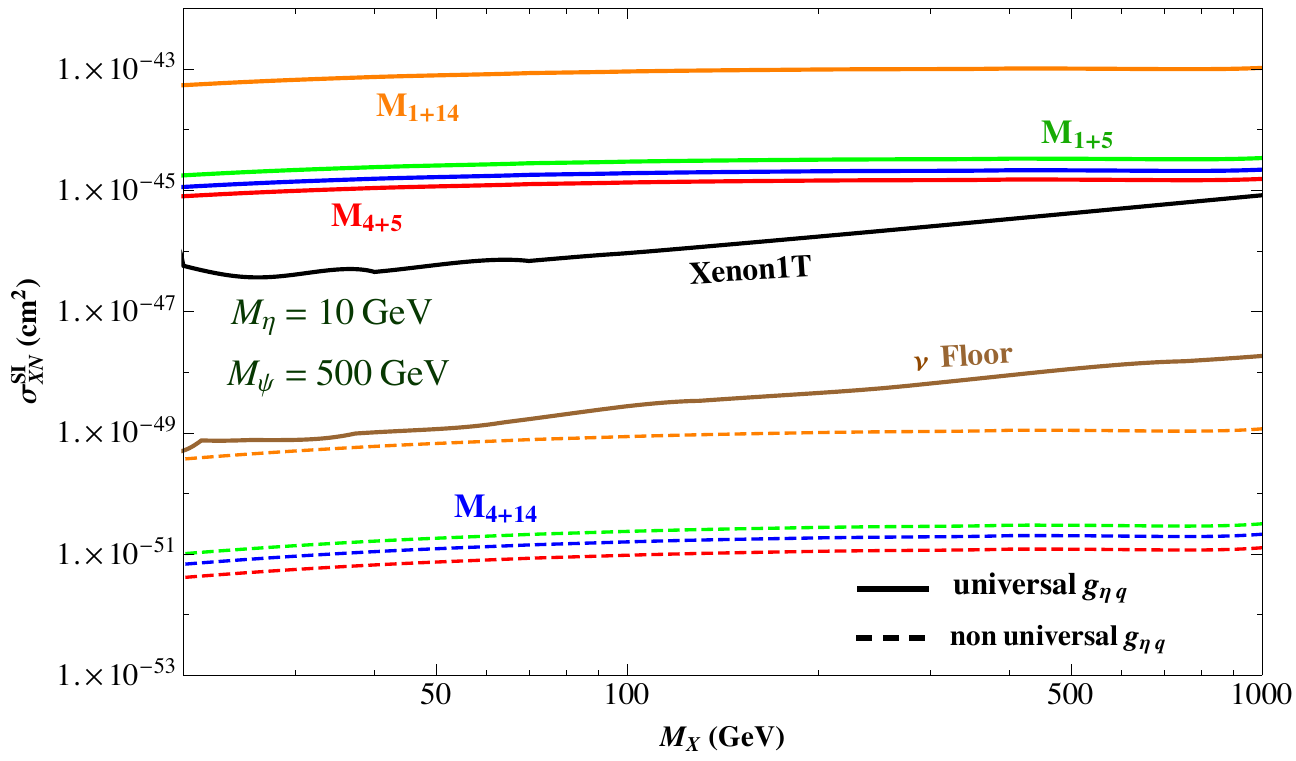}
\caption{\sf Implementing the couplings of~\eqref{Couplings-fRfReta} with the universal assumption $\operatorname{Im}(y_{q \psi })=\operatorname{Im}(y_{u \psi })=\operatorname{Re}(y_{q \psi })=\operatorname{Im}(y_{q u})=1/2$ and $\operatorname{Im}(\tilde{y}_{u \psi })=\operatorname{Im}(\tilde{y}_{q \psi })=1/2$ (thick curves), compared to the non-universal situation with couplings proportional to the quark mass (dashed), for $M_\eta=10$ GeV and $M_\Psi=500$ GeV.}
\label{Cross-sections-uni-non-uni-couplings}
\end{center}
\end{figure}

When implementing the effective couplings of~\eqref{Couplings-fRfReta}, the Xenon1T limit and the neutrino floor can be partially or totally evaded. For a light mediator mass of 10 GeV (Fig.~\ref{Cross-sections} upper plot), Xenon1T excludes all the models for a top partner mass scale of $M_\Psi=500$ GeV (thick curves). A heavier scale of 1500 GeV (dashed curves) suppresses the cross sections in all the models, partially eluding the Xenon1T limits excepting in $\oB$ (orange). We have set\footnote{A preciser evaluation would imply  the non-universal assumption $\operatorname{Im}(y_{q \psi })=\operatorname{Im}(y_{u \psi })=\operatorname{Re}(y_{q \psi })=\operatorname{Im}(y_{q u})=m_q/v$ and $\operatorname{Im}(\tilde{y}_{u \psi })=\operatorname{Im}(\tilde{y}_{q \psi })=m_q/v$.  This would lead to set $0.0055$, $0.017$ and $0.69$ correspondingly for $q=c,b,t$.  So, in the light of these values it was chosen 1/2 for the previous couplings, overestimating then the charm and bottom contributions, and underestimating the top one. The non-universal assumption is implemented later.
} $\operatorname{Im}(y_{q \psi })=\operatorname{Im}(y_{u \psi })=\operatorname{Re}(y_{q \psi })=\operatorname{Im}(y_{q u})=1/2$ and $\operatorname{Im}(\tilde{y}_{u \psi })=\operatorname{Im}(\tilde{y}_{q \psi })=1/2$. Additionally, the Yukawa couplings are chosen as $y _R=\tilde{y}_R=1$, while $y _L$ and $\tilde{y}_L$ are properly fixed in order to maintain the top mass at its experimental value through its corresponding mass formula. This is obtained for the $\bf{5}$-plet and $\bf{14}$-plet scenarios in~\cite{Yepes:2018dlw,Norero:2018dfg,Yepes:2017pjr}.

It is remarkable to observe how for a mediator mass of 50 GeV (Fig.~\ref{Cross-sections} centred plot), the Xenon1T limit, and even the neutrino floor, are evaded. Indeed, for $M_\Psi=500$ GeV all the models are totally below Xenon1T, with the fourplet models and $\oA$ (red-blue-green) nearby the neutrino floor, and even eluding it for a DM mass higher than~$\sim 200$-$300$ GeV. Increasing the scale up to $M_\Psi=1500$ GeV, the latter models completely elude the neutrino floor in all the DM mass range 10-1000 GeV. 

The previous suppression is further enhanced for a higher mediator mass of 100 GeV (Fig.~\ref{Cross-sections} lower plot). Indeed, for $M_\Psi=500$ GeV, the Xenon1T limit is notoriously evaded along the entire DM mass range in the fourplet scenarios and $\oA$. Through the same DM mass region, such models are remarkably below the neutrino floor for a top partner mass scale of 1500 GeV. In this case, $\oB$ becomes completely below Xenon1T, partially eluding the neutrino floor for a DM mass heavier than~$\sim 200$ GeV.

The non-universal assumption $\operatorname{Im}(y_{q \psi })=\operatorname{Im}(y_{u \psi })=\operatorname{Re}(y_{q \psi })=\operatorname{Im}(y_{q u})=m_q/v$ and $\operatorname{Im}(\tilde{y}_{u \psi })=\operatorname{Im}(\tilde{y}_{q \psi })=m_q/v$ (consistent with MFV), with $q=c,b,t$, is implemented in Fig.~\ref{Cross-sections-uni-non-uni-couplings}. As it can be seen, the proportionality of the couplings to the quark mass induces an additional suppression on the SI cross sections with respect to the universal assumption considered in Fig.~\ref{Cross-sections}. Notice how for a relatively small mediator mass of 10 GeV the SI cross section can be suppressed below the Xenon1T limit and the neutrino floor as well for $M_\Psi=500$ GeV. This suggests that the DD upper bounds can be totally evaded even for a mediator masses lighter than~$\sim 10$ GeV when invoking the non-universal assumption. This evasion would be further enhanced for a higher top partner scales.

The cross sections displayed in Fig.~\ref{Cross-sections} were computed varying $\xi$, \ie varying $M_\Psi$ for a fixed $g_\Psi$. From~\eqref{Amplitude-cross-section}-\eqref{final-Amplitude} and the $\xi$-dependence in~\eqref{Couplings-fRfReta}, we roughly have 
\be
\sigma_{{\rm DM} N}^{\rm SI}\approx g^4_{\eta q}\sim \xi^2\sim \frac{g^4_\Psi\,v^4}{M^4_\Psi}.
\label{Estimation}
\ee

\begin{figure}
\begin{center}
\includegraphics[scale=0.355]{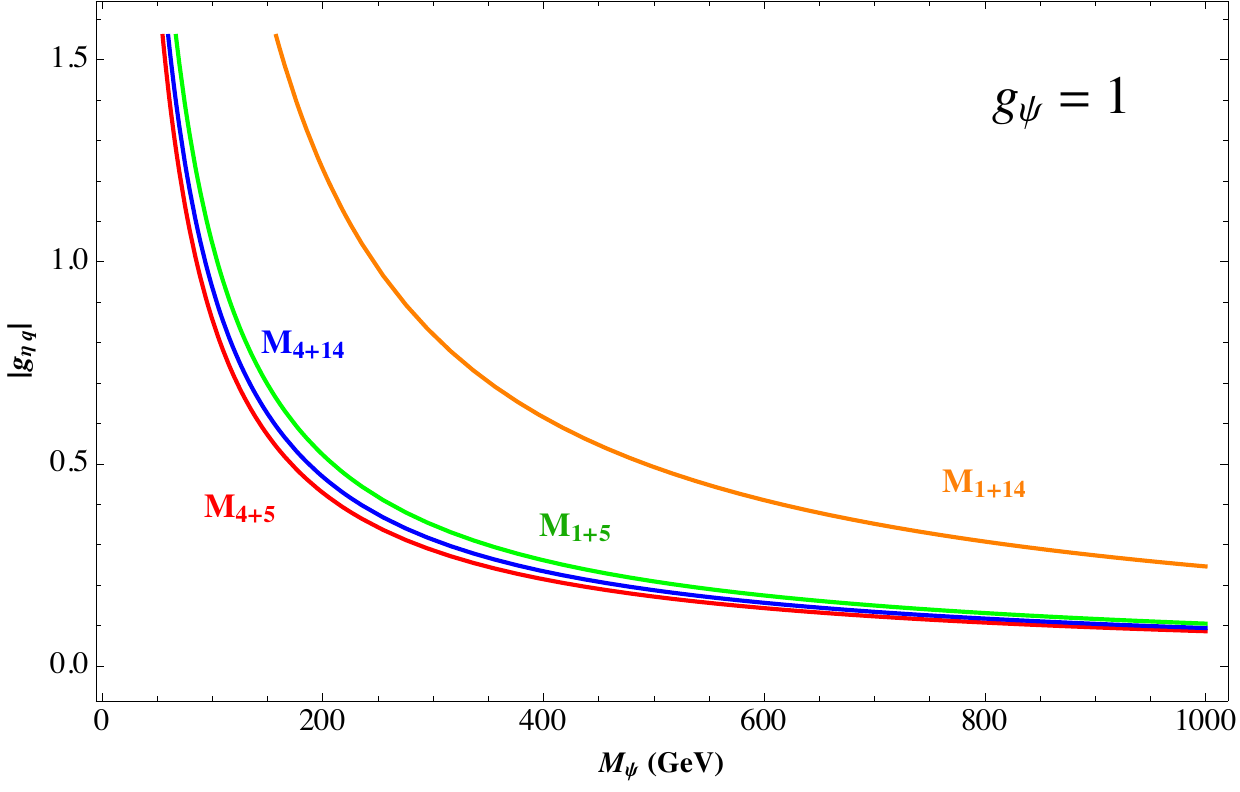}
\caption{\sf Dependence of the absolute value of the couplings $g_{\eta q}$ with the top partner mass scale  $M_\Psi$, for a fixed $g_\Psi=1$. The couplings $g_{\eta q}$ are independent of the DM mass $M_X$ and the mediator mass $M_\eta$ (see Eqs.~\eqref{Couplings-fRfReta} and~\eqref{eta-parameters}). The universal assumption for the involved couplings is implemented.}
\label{Couplings-vs-Mpartner}
\end{center}
\end{figure}

\nt If $M_\Psi$ augments with a fixed $g_\Psi$, then a suppressing factor is introduced in addition to the one encoded by the $\eta$ and $\tilde{\eta}$-dependence in~\eqref{Couplings-fRfReta}. This can be visualized in Fig.~\ref{Couplings-vs-Mpartner} for a fixed\footnote{In Fig.~\ref{Cross-sections} we have explored two situations $M_\Psi=500,\,1500$ GeV for a fixed $g_\Psi=1$, which corresponds to $\xi\sim 0.24,\,0.027$. The former value is compatible with the latter EWPT bounds, as well as the limits from the vector resonance direct production and the expected single Higgs production at the LHC. The latter value is favoured by the 95\% combined limit from direct production of vector resonances at the LHC~\cite{Contino:2013gna} and for a mass of $\sim 2$ TeV.} $g_\Psi=1$, whose implied behaviour\footnote{The couplings $g_{\eta q}$ do not dependent on the DM and mediator masses $M_X$ and $M_\eta$, as it can be seen from  Eqs.~\eqref{Couplings-fRfReta} and~\eqref{eta-parameters}. They only depend on the top partner mass scale $M_\Psi$ as the effective couplings  are the outcome of integrating out the top partners from the physical spectrum. The dependence on $M_X$ and $M_\eta$ appears in the cross section via the loop function (see~\eqref{final-Amplitude}).} favours the evasion of the DD experiments, as well as the neutrino floor in some DM mass regions. In contrast, this situation differs with respect to the scenario of $M_\Psi$ and $g_\Psi$ simultaneously changing (fixed $\xi$). The latter situation is explicitly depicted in Fig.~\ref{Cross-sections-fixed-xi} App.~\ref{fixed-xi}. 

\begin{figure}
\begin{center}
\includegraphics[height=3cm, width=4cm]{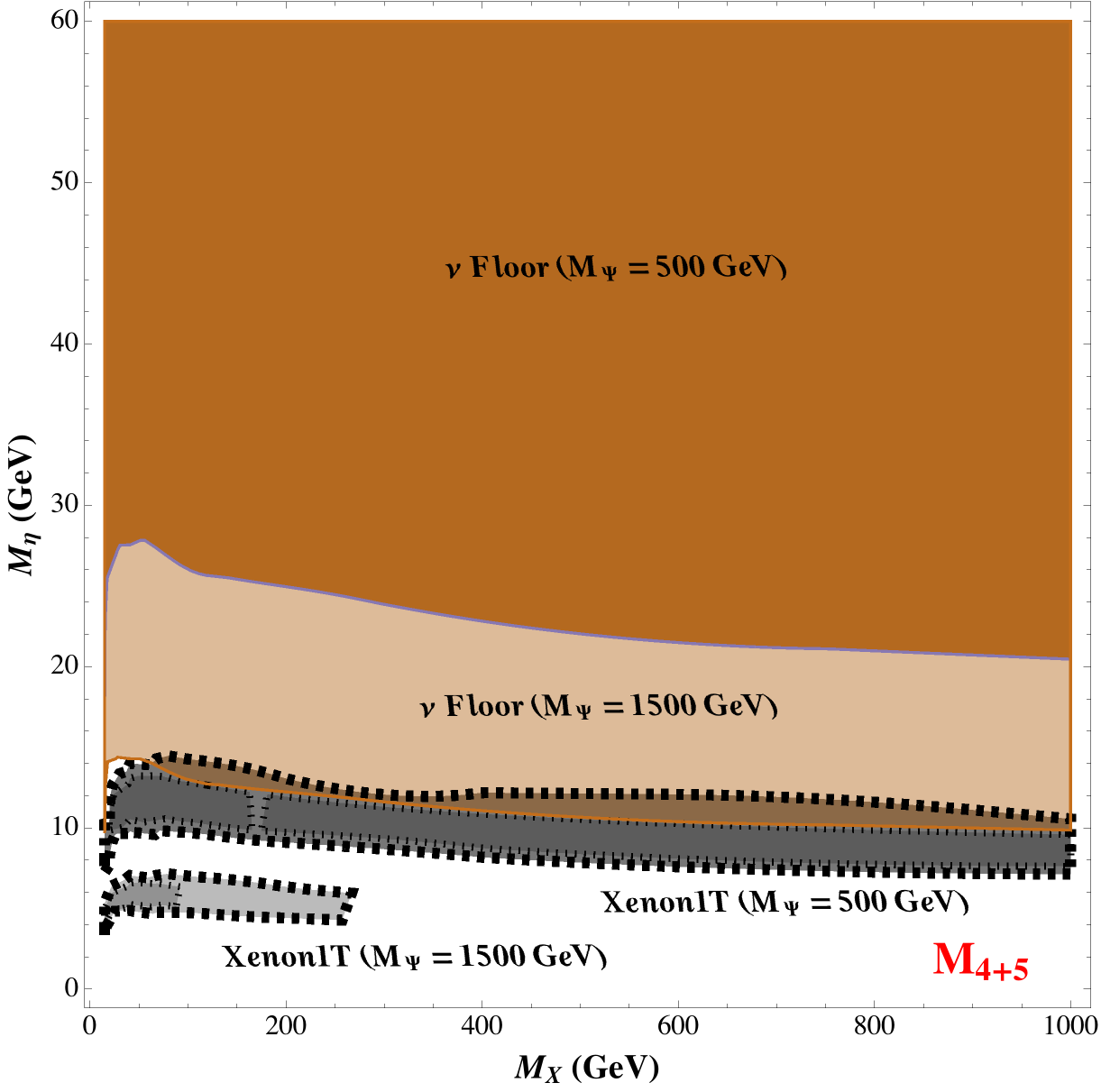}
\hspace*{0.4cm}
\includegraphics[height=3cm, width=4cm]{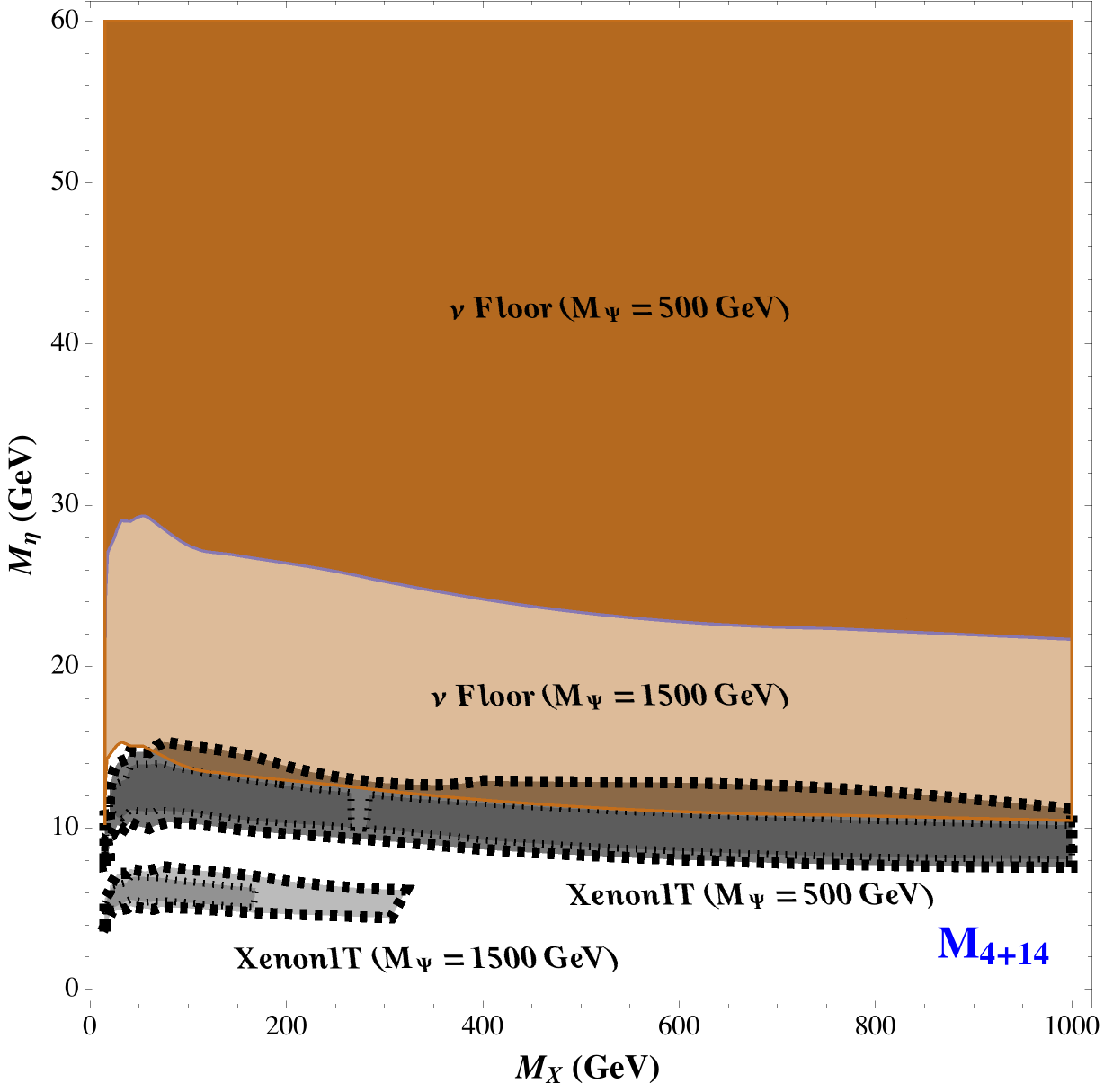}\\
\vspace*{0.5cm}
\includegraphics[height=3cm, width=4cm]{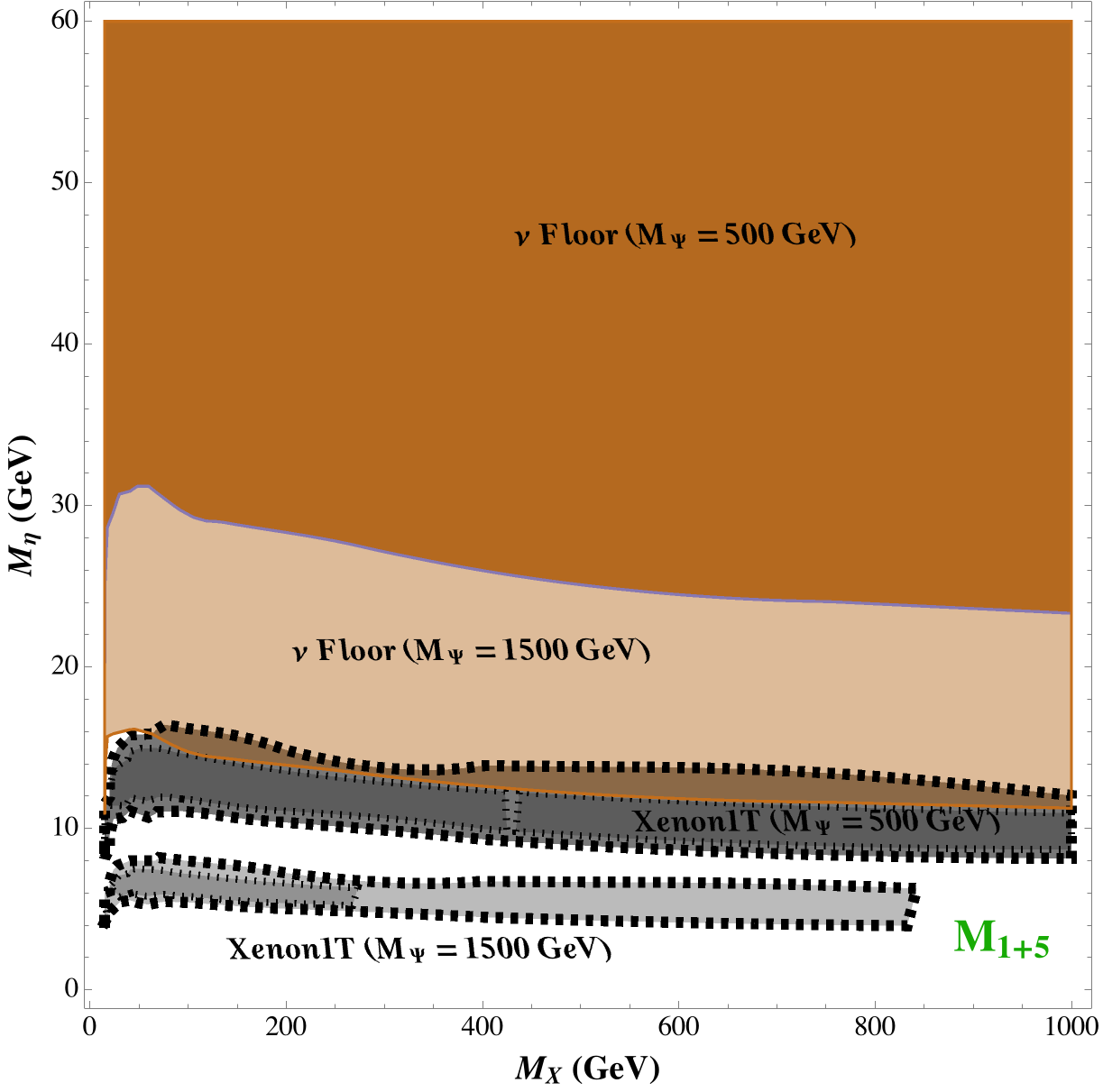}
\hspace*{0.4cm}
\includegraphics[height=3cm, width=4cm]{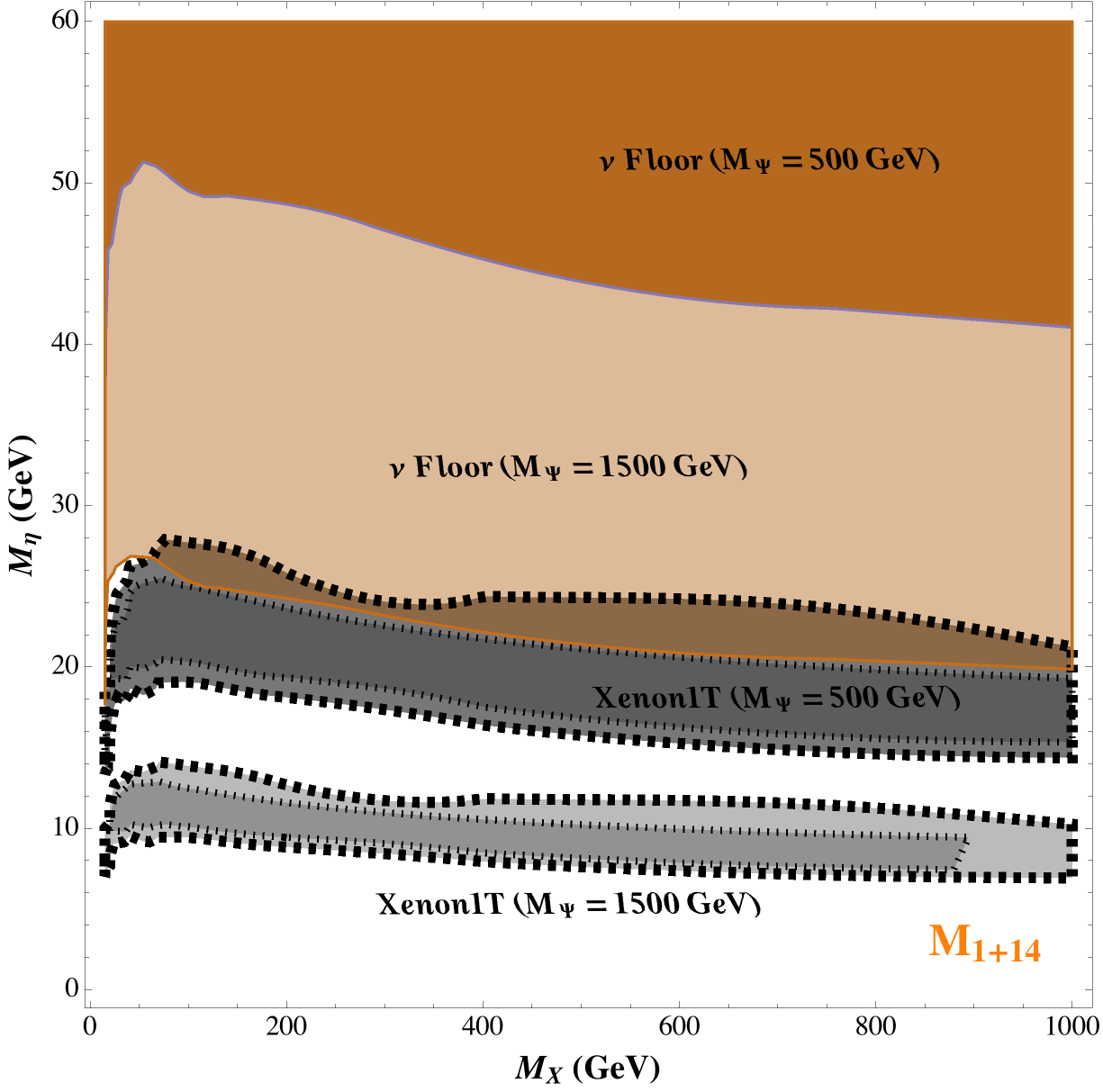}
\caption{\sf Parameter spaces $(M_\Psi,\,M_\eta)$ at the fourplet and singlet scenarios (top-bottom) allowed by the Xenon1T upper limits (gray areas) and the neutrino floor (orange). Darker-lighter coloured regions correspond to $M_\Psi = 500,\,1500$ GeV. Dotted-dashed bordered areas stand for the $1\sigma$-$2\sigma$ Xenon1T values.}
\label{Parameter-spaces}
\end{center}
\end{figure}

The parameter space for the DM-mediator masses that are permitted by Xenon1T (gray), and the regions with the cross section below the neutrino floor (orange) are shown in Fig.~\ref{Parameter-spaces}, for the fourplet and singlet models (upper-lower plots). The 2-loop contribution has not been included as its contribution is less relevant than the triangle and box diagrams. This is explicitly justified in Fig.~\ref{Box-Triangle-2-loop}.

Notice that mediator masses in the range $\sim 10$-15 GeV are constrained by Xenon1T for the fourplet models together with $\oA$, while $\sim 15$-25 GeV in $\oB$ for $M_\Psi=500$ GeV (darker areas). In this case, the neutrino floor becomes eluded for a mediator mass roughly heavier than~$\sim 25$ GeV in the whole DM mass regions for the fourplet models and $\oA$, whereas  mediator masses higher than~$\sim 45$ GeV in $\oB$. As soon as the top partner is heavier as 1500 GeV (lighter areas), much smaller mediator masses in the range $\sim 4$-8 GeV are constrained by Xenon1T in the fourplet scenarios (for the DM mass range $\sim 10$-230 GeV) and $\oA$ (for the broader DM mass region $\sim 10$-820 GeV). Mediator masses as $\sim 8$-14 GeV are allowed in $\oB$ throughout the DM mass range. The neutrino floor becomes eluded for mediator masses above~$\sim 12$ GeV  in all the explored DM mass region for the fourplet models and $\oA$. Above~$\sim 22$ GeV for mediator masses in $\oB$ it is permitted to be below the neutrino floor. 

All in all, the fourplet scenarios and $\oA$ better suppress the WIMP-nucleon cross sections rather than $\oB$. Its coupling $g_{\eta q}$ in~\eqref{Couplings-fRfReta} involves a suppression only from the $\xi$-dependence, unlike to the others whose suppression is directly augmented by the contribution of the parameters $\eta$ and $\tilde{\eta}$. This feature disfavours $\oB$ in effectively eluding the latest Xenon1T bounds. Future observations will help us in discriminating the best framework among the fourplet models and $\oA$ for the explanation of the DD experiments.


\section{Summary}
\label{Summary}

\nt The WIMP-nucleon scattering cross section in a simple vector dark matter model is analysed at the loop-level. A scalar mediator is coupled only to the vector dark field and to the SM quarks via pseudo-bilinear interactions. The latest direct detection experiments are imposed to bound the involved parameter space for the dark matter and the mediator masses. We provide effective mediator-quark couplings whose effect suppresses the total WIMP-nuclei cross section, evading thus the recent direct detection Xenon1T limits. We motivate such suppression from New Physics scales reachable at the LHC and inspired by Composite Higgs scenarios. They naturally source the top partners, whose implied masses suppress the effective pseudo-bilinear couplings. In summary, we are able to evade the upper current experimental bounds, predicting also reasonable regions for the dark matter-mediator masses, and experimentally consistent in a coherent theoretical framework.

%
%
%
\section*{Acknowledgments}
\nt The author acknowledges the valuable data from Xenon1T and the neutrino floor provided by Daniel Coderre, as well as interesting discussions with Felipe J. Llanes-Estrada, Juan J. Sanz-Cillero and M. Cermeño. The author also acknowledges the hospitality received at the Universidad Complutense de Madrid, where part of this work has been carried out. The author finally acknowledges comments from Alfonso Zerwekh and Jose D. Ruiz Álvarez. J.Y. thanks  the support of Fondecyt (Chile) grant No. 3170480.

\appendix
\small

\section{CCWZ formalism}
\label{CCWZ}

\nt The $SO(4) \simeq SU(2)_L \times SU(2)_R$ unbroken generators and the broken ones parametrising the coset $\textrm{SO}(5)/SO(4)$ in the fundamental representation are
\begin{align}
(T^a_\chi)_{IJ} &= -\frac{i}{2}\left[\frac{1}{2}\varepsilon^{abc}
\left(\delta_I^b \delta_J^c - \delta_J^b \delta_I^c\right) \pm
\left(\delta_I^a \delta_J^4 - \delta_J^a \delta_I^4\right)\right],\\
T^{i}_{IJ} &= -\frac{i}{\sqrt{2}}\left(\delta_I^{i} \delta_J^5 - \delta_J^{i} \delta_I^5\right)\,,
\label{eq:SO4_gen-SO5/SO4_gen}
\end{align}

\nt with $\chi= L,\,R$, and $a= 1,2,3$, while $i = 1, \ldots, 4$. The normalization of $T^{A}$'s is chosen as ${\rm Tr}[T^A, T^B] = \delta^{AB}$.

\section{Top partners EOM}
\label{Partners-EOM}

\nt For a heavy top partners mass scenario the corresponding fields may be integrated out from the physical spectrum, via their associated equations of motion. Concerning the models $\fA$ and $\oA$ altogether, and after diagonalising the mass terms at $\LL_{\text{mix}}$ in~\eqref{fA-oA-mix}, plus the terms associated to their kinetic composite sector, we obtain the field redefinitions for the left handed components as
\beq
\begin{aligned}
&\mathcal{T}_L\to -\frac{\xi}{4}\,\frac{\eta_L}{\eta _L^2+1}\, t_L,\,\quad
&&\tilde{\mathcal{T}}_L\to \sqrt{\frac{\xi}{2}}\,\frac{ \tilde{\eta }_L}{\left(\tilde{\eta }_R^2+1\right)\sqrt{\eta _L^2+1}}\,t_L,\,\\[4mm]
&X_{\text{2/3},L}\to \frac{\xi}{4}\,\frac{\eta _L}{\sqrt{\eta _L^2+1}}\,t_L ,\,
&&\mathcal{B}_L\to 0,\,
\end{aligned}
\label{EOM-fA-oA-L}
\eeq

\nt while for the right handed components one obtains
\beq
\begin{aligned}
&\mathcal{T}_R\to \sqrt{\frac{\xi}{2}}\,\frac{\eta _R }{\left(\eta _L^2+1\right) \sqrt{\tilde{\eta }_R^2+1}}\,t_R,\,
&&\quad\tilde{\mathcal{T}}_R\to -\frac{\xi}{2}\,\frac{\tilde{\eta }_R}{\tilde{\eta }_R^2+1}\,t_R,\,\\[4mm]
&X_{\text{2/3},R}\to -\sqrt{\frac{\xi}{2}}\,\frac{\eta _R }{\sqrt{\tilde{\eta }_R^2+1}}\,t_R,\,
&&\quad \mathcal{B}_R\to -\tilde{\eta }_R\,b_R,
\end{aligned}
\label{EOM-fA-oA-R}
\eeq

\nt where the parameters $\eta_{L(R)}$ are defined through

\be
\eta _{L(R)}\equiv \frac{y_{L(R)} \mathit{f}}{M_{\bf{4}}},\qquad\qquad
\tilde{\eta}_{L(R)}\equiv \frac{\tilde{y}_{L(R)} \mathit{f}}{M_{\bf{1}}}.
\label{eta-parameters}
\ee

\nt Likewise, from $\LL'_{\text{mix}}$ the in~\eqref{fB-oB-mix}, we obtain for $\fB$ and $\oB$

\beq
\begin{aligned}
&\mathcal{T}_L\to -\frac{5\,\xi}{4}\,\frac{\eta_L}{\eta _L^2+1}\, t_L,\,
&&\quad\tilde{\mathcal{T}}_L\to -\frac{\sqrt{2\,\xi }}{\sqrt{\eta _L^2+1}}\tilde{\eta }_L\,t_L,\,  \\[4mm]
& X_{\text{2/3},L}\to \frac{3\,\xi}{4}\,\frac{\eta _L}{ \sqrt{\eta _L^2+1}}\,t_L,\,
&& \quad\mathcal{B}_L\to -\frac{\xi}{2}\,\frac{\eta_L}{\eta _L^2+1}\,b_L ,\,
\end{aligned}
\label{EOM-fB-oB-L}
\eeq

\nt while for the right handed components one obtains
\beq
\begin{aligned}
&\mathcal{T}_R\to -\sqrt{2\,\xi }\frac{\sqrt{\tilde{\eta }_R^2+1}}{\eta _L^2+1}\eta _L\eta _R t_R,\,
&&\quad \tilde{\mathcal{T}}_R\to -\tilde{\eta }_R\,t_R,
\\[4mm]
&\mathcal{B}_R\to -\tilde{\eta }_R\,b_R,
&& \quad X_{\text{2/3},R}\to 0
\end{aligned}
\label{EOM-fB-oB-R}
\eeq

\nt with the coefficients $\eta$ similarly defined as in~\eqref{eta-parameters}, but with the Yukawa couplings $y$ and $\tilde{y}$ suitably replaced by those in $\fB$ and $\oB$. Both the left and right handed components for the field $X_{\text{5/3}}$ are zero in all the models.

\begin{figure}
\begin{center}
\includegraphics[height=3cm, width=4cm]{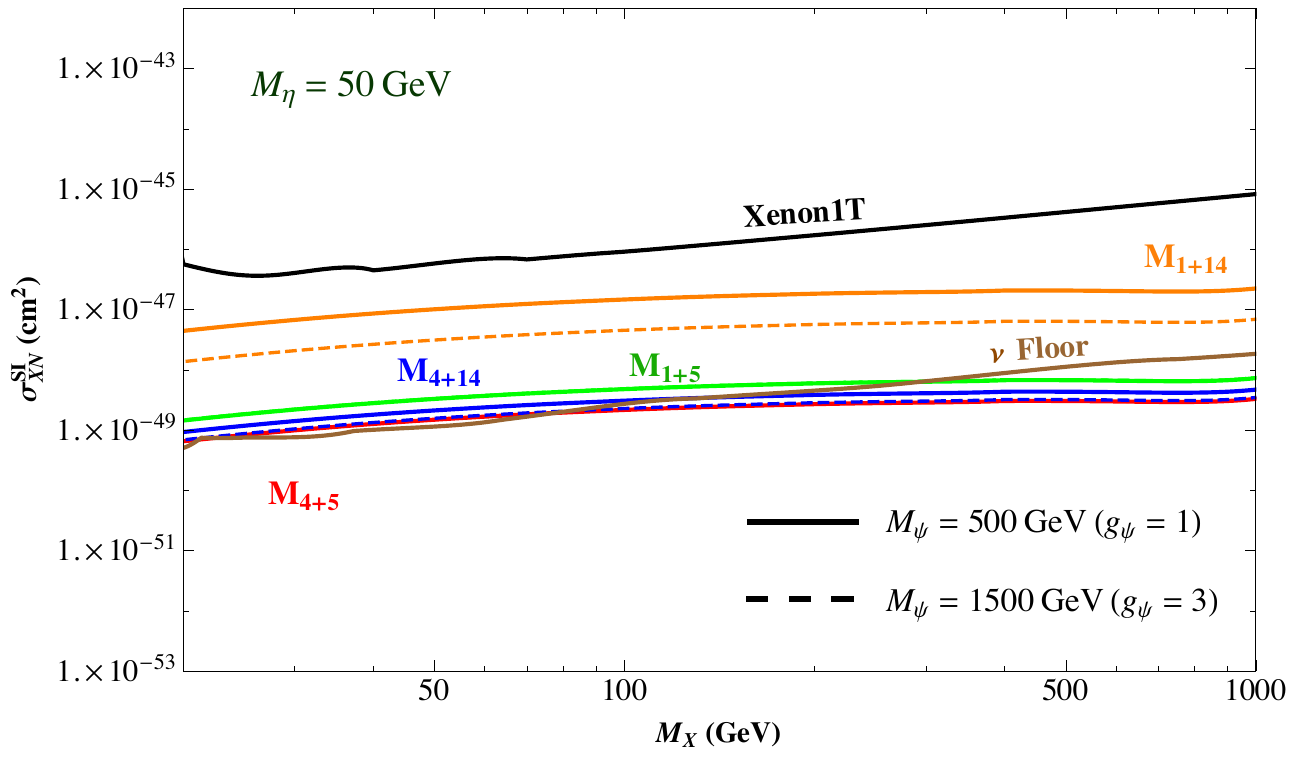}
\hspace*{0.25cm}
\includegraphics[height=3cm, width=4cm]{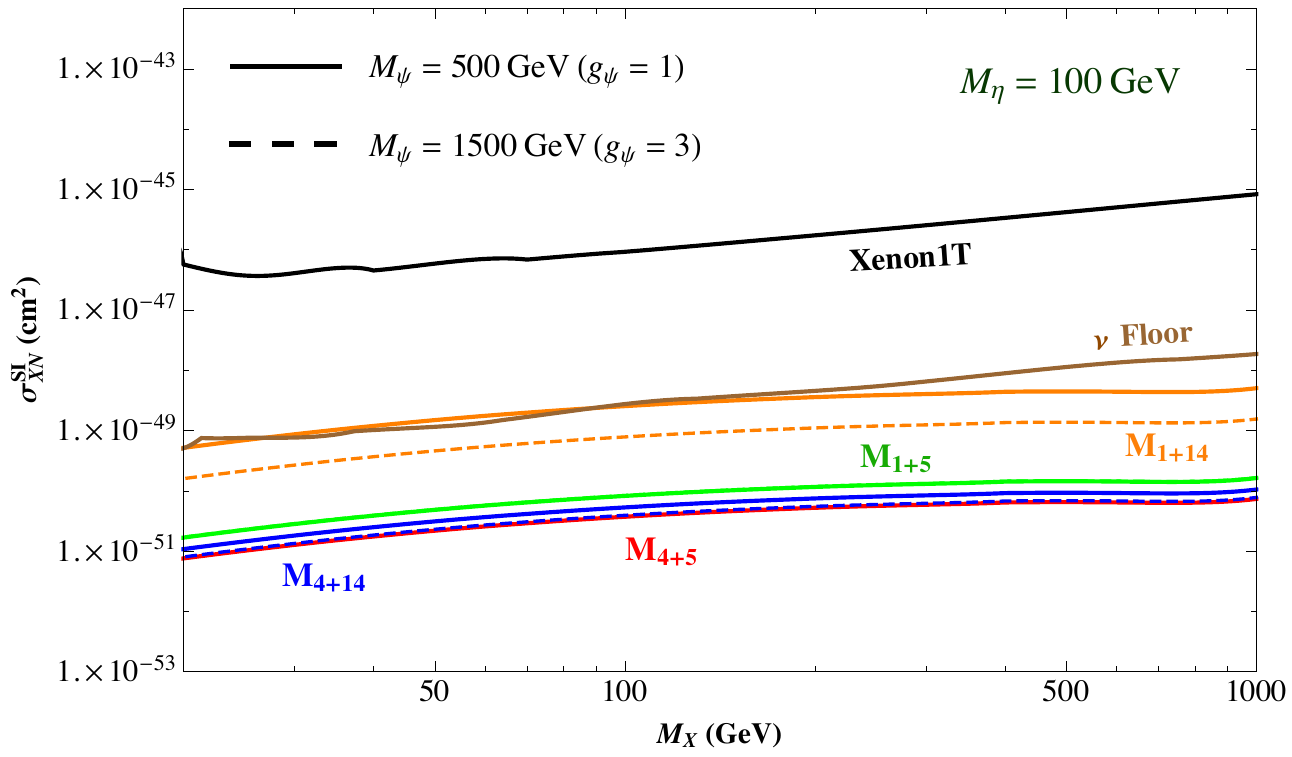}
\caption{\sf SI DM-nucleon scattering cross section for a fixed $\xi$. We have explored for a mediator mass $M_\eta=50,\,100$ GeV (left-right plot), two different cases:  $M_\Psi=500,\,1500$ GeV (thick-dashed curves) with associated couplings $g_\Psi=1,\,3$, the which correspond to $\xi\sim 0.2$. The Xenon1T limit and the neutrino floor (black and brown), are further evaded for heavier mediator masses rather than for heavier top partner masses.}
\label{Cross-sections-fixed-xi}
\end{center}
\end{figure}
\section{WIMP-nucleon cross sections at fixed $\xi$}
\label{fixed-xi}

\nt For a fixed $\xi$ value, \ie  $M_\Psi$ and $g_\Psi$ properly varying, the effective pseudo-bilinear couplings in~\eqref{Couplings-fRfReta} will depend on the top partner mass $M_\Psi$ only via the parameters $\eta_{L(R)}$ and $\tilde{\eta}_{L(R)}$. In this case, the induced suppression in the cross sections is less notorious as in the case of a decreasing $\xi$, \ie an increasing $M_\Psi$ with fixed $g_\Psi$. 

Fig.~\ref{Cross-sections-fixed-xi} shows the SI DM-nucleon scattering cross section for mediator masses $M_\eta=50,\,100$ GeV (left-right plots) and $M_\Psi=500,\,1500$ GeV (thick-dashed curves), with the corresponding couplings $g_\Psi=1,\,3$. The parameter $\xi$ remains fixed then at a rough value of 0.2. 

Notice how the cross sections suffer a less suppression when augmenting the top partner mass, compared with those in Fig.~\ref{Cross-sections}. The evasion of Xenon1T and the neutrino floor mainly depends on the mediator mass in this case.

\section{Triangle diagram contribution}
\label{Triangle-diagram}

\nt A triangle diagram contributing to the SI cross section emerges once a coupling $\eta$-$\eta$-$h$ is accounted for in this CHM set-up. To do so, a calculable Higgs potential is in order. However, this minimal CHM does not have the sufficient structure (states and couplings) to make it calculable. In principle this would be possible by enlarging the initial $SO(5)/SO(4)$ coset assumed here to $SO(5)_L\times SO(5)_R$ (as in a two-site model~\cite{Panico:2011pw,Panico:2015jxa}) spontaneously broken to ${SO(5)_V}$. Now the coset $SO(5)_L\times SO(5)_R/SO(5)_V$ would be parametrised by
\begin{equation}
\displaystyle
U\, =\, \exp\left[i \frac{\sqrt{2}}{f}\Pi_A T^A \right], 
\label{GB-matrix-larger-symmetry}
\end{equation}

\nt with  $T^A=\left\{T^a_{L(R)},\,\,T^{\widehat{a}}\right\}$ the unbroken ($a=1,2,3$), and  broken generators ($\widehat{a}=1,2,3,4$). There are ten Goldstones $\Pi_A$, four of them forming a fourplet (identified with the Higgs field). The remaining six in the adjoint of $SO(4)$ are removed by gauging. In addition, it is possible to introduce the spurions ${\mathcal G}^\alpha = g_0 T^\alpha_L$ and ${\mathcal G}' = g'_0 T^3_R$  associated to the gauging of the $SU(2)_L$ and $U(1)_Y$ subgroups of $SO(5)$, while $\widetilde {\mathcal G}_a = \widetilde g_\rho T^a$ for the gauging of $SO(4)$, subgroup of $SO(5)_R$. 

With these spurions and their transformation properties under the total symmetry, it is possible to construct the leading contributions to the potential from the operators $\textrm{Tr} \left[{\mathcal G}_\alpha{\mathcal G}_\alpha U \widetilde{\mathcal G}_a\widetilde{\mathcal G}_aU^t\right]$ and $\textrm{Tr}\left[{\mathcal G}'{\mathcal G}' U \widetilde{\mathcal G}_a\widetilde{\mathcal G}_a U^t\right]$. The Higgs potential turns out to be logarithmically divergent~\cite{Panico:2011pw,Panico:2015jxa}. Analogous operators coupled to the mediator are straightforward
\begin{equation}
\frac{c_{g\eta}}{16 \pi^2} f^3 \eta\Tr\left[{\mathcal G}_\alpha {\mathcal G}_\alpha U
\widetilde {\mathcal G}_\alpha \widetilde {\mathcal G}_\alpha U^T\right],
\quad
\frac{c_{g'\eta}}{16 \pi^2} f^3 \eta\Tr\left[{\mathcal G}' {\mathcal G}' U
\widetilde {\mathcal G}_\alpha \widetilde {\mathcal G}_\alpha U^T\right]
\label{eta-Higgs-ops}
\end{equation}

\nt They give rise to the interactions $\eta$-$h$  and $\eta$-$h$-$h$. After diagonalising the former, the latter finally leads to the effective interaction $\eta$-$\eta$-$h$, encoded by the coupling $g_{\eta\eta h}(c_{g\eta},c_{g'\eta},\theta)$. It depends on a linear combination of $c_{g\eta}$, $c_{g'\eta}$ and on the involved mixing angle
\begin{equation}
\tan 2\theta \sim  \frac{c_{g\eta} + c_{g'\eta}}{16 \pi^2} \frac{ f^2}{M^2_h-M^2_\eta}.
\label{Mixing-angle}
\end{equation}

\nt Relative factors coming from the traces in~\eqref{eta-Higgs-ops} are omitted as we intend to have just a quick glance of the computation. The WIMP-quark scattering amplitudes from the triangle, box and the two-loop diagram will conveniently depend on the DM-mediator, fermion-Higgs and fermion-mediator couplings after scalar diagonalisation, \ie $g_{X\eta}(\theta)$, $g_{qqh}(\theta)$ and $g_{\eta\eta h}(c_{g\eta},c_{g'\eta},\theta)$.
\begin{figure}
\begin{center}
\includegraphics[height=3cm, width=4cm]{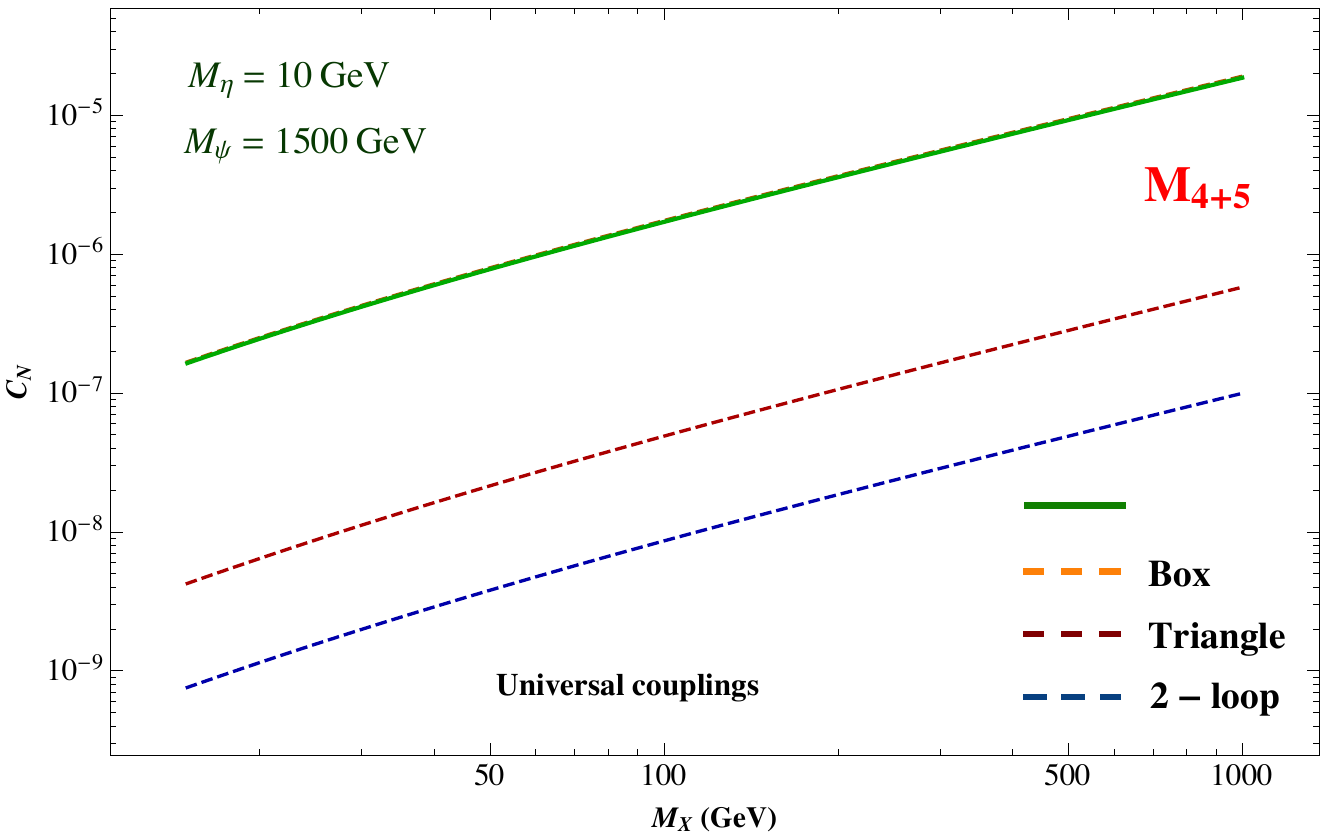}
\includegraphics[height=3cm, width=4.5cm]{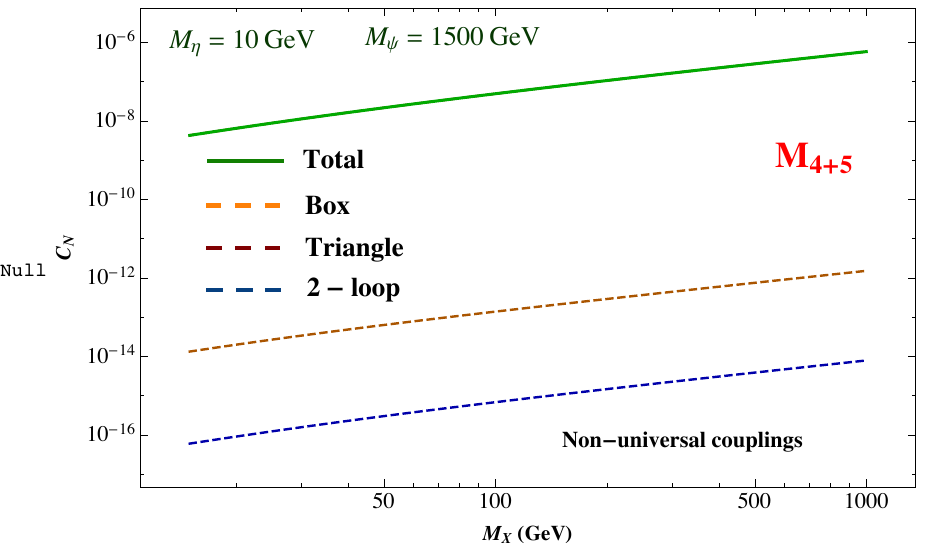}
\caption{\sf Left: combined effect of the box (orange), triangle (red) and the two-loop diagrams (blue) on the  total loop function (green thick) at the model $\fA$, for $M_\eta=10$ GeV and $M_\Psi=1500$ GeV with universal couplings.  Right: same plot with non-universal couplings.}
\label{Box-Triangle-2-loop}
\end{center}
\end{figure}

This result is basically depending on the strength of the weighting coefficients of~\eqref{eta-Higgs-ops}, signaling thus the strong dependence of the cross section on the larger symmetry group underlying our framework. As an example, Fig.~\ref{Box-Triangle-2-loop} displays the combined effect on the whole loop function $\cC_N$ of~\eqref{final-Amplitude} (green thick) in $\fA$ for the universal (left plot) and non-universal (right plot), from the box (orange), triangle (red) and the 2-loop diagram (blue), with $M_\eta=10$ GeV and $M_\Psi=1500$ GeV. It has been set $c_{g\eta},\,c_{g'\eta}\sim 1$, assuming also the relative factors from the traces in~\eqref{eta-Higgs-ops} to be order one. For the non-universal couplings the triangle loop is more relevant than the box contribution, as the fermion-mediator couplings enter quadratically  at the former while in a quartic dependence at the latter. Two loops as it can be seen, are always the less relevant contribution.

A more refined study with precise relative factors from the operators in the scalar potential is in order. Also, a complete analysis of the involved parameter spaces for the operator coefficients $c_{g\eta}$ and $c_{g'\eta}$ must be done, together with the impact of the triangle diagram contribution on the parameter spaces $(M_\Psi,\,M_\eta)$ of Fig.~\ref{Parameter-spaces}. All those issues have been postponed for a future work.

\section{Loop functions}
\label{Loop-function}

\nt The loop function $\cC_N$ in~\eqref{Amplitude-cross-section} encodes contributions in~\eqref{final-Amplitude}-\eqref{final-Wilson} from the triangle and box diagrams in Fig.~\ref{Box-diagrams}-\ref{Triangle-Two-Loop-diagrams}. They are reported here in terms of the Passarino-Veltman functions~\cite{Passarino:1978jh}. We obtain
\begin{align}
& f^{\text{tri}}_q= g^2_{X\eta}\,g_{\eta\eta h}\,\frac{M^2_X}{M^2_h\,v}\,\frac{1}{(4\pi)^2}\, C_0\left(M^2_X,\, M^2_\eta,\,M^2_X\right),\\[6mm] \nn 
& f^{\text{box}}_q= -4\,g^2_{X\eta}\,g^2_{\eta q}\,M^2_X\,\frac{1}{(4\pi)^2}\,\times\\ \nn
&\phantom{f^{\text{box}}_q=}\left[ Z_{00}\left(M^2_X,\,M^2_X ,\,M^2_\eta\right)\,+\,\frac{M^2_X}{4}Z_{11}\left(M^2_X,\,M^2_X ,\,M^2_\eta\right)\right],
\end{align}

\begin{align}
& g^{\text{box}}_q= -4\,g^2_{X\eta}\,g^2_{\eta q}\,M^4_X\,\frac{1}{(4\pi)^2}\,Z_{11}\left(M^2_X,\,M^2_X ,\,M^2_\eta\right),
\\[6mm]
& f^{\text{tri}}_G= \sum_{q=c,b,t} f^{\text{tri}}_q = 3\,f^{\text{tri}}_q,\\[6mm]
& f^{\text{box}}_G = -\frac{1}{(4\pi)^2}\sum_{q=c,b,t} g^2_{X\eta}\,g^2_{\eta q}\,M^2_X\,F_X\left(M^2_X,\,M^2_X ,\,M^2_\eta,\,m^2_q\right),
\end{align}

\nt with the Passarino-Veltman function $C_0$ as
\be
\int \frac{\text{d}^{4}k}{(2\pi)^4} \frac{1}{[(p+k)^2 - M^2]\, [k^2 - m^2]^2} = \frac{i}{(4\pi)^2}\, C_0(p^2,\, m^2,\,M^2)
\ee

\nt and the $Z$ functions extracted from
\begin{align}
& \int \frac{\text{d}^{4}k}{(2\pi)^4} \frac{k^\mu k^\nu}{[(p+k)^2 - M^2]\,k^4\,[k^2 - m^2]^2} \\[5mm] \nn
&\phantom{\int \frac{1}{2}} = \frac{i}{(4\pi)^2}\Big[p^\mu p^\nu\,Z_{11}(p^2,\,M^2,\,m^2) + g^{\mu \nu}\,Z_{00}(p^2,\,M^2,\,m^2)  \Big].
\end{align}

\nt For the two-loop computation it is followed a similar procedure described in~\cite{Ertas:2019dew}. Firstly, the calculation of the leading order effective mediator-gluon vertexes is done via the Fock-Schwinger gauge, in which the gluon field is directly written in terms of the field strength~\cite{Hisano:2010ct,Novikov:1983gd}. Afterwards, such vertexes are included into the remaining triangle diagram of the full two-loop, obtaining the function $F_X$ 
\begin{align}
& F_X= \int^1_0 dx\left[(1-x)x+\frac{3}{2}\left(x^2+(1-x)^2\right)\right]\times \nn \\[2mm] 
&\phantom{F_X= \int^1_0 dx}\times\left[\frac{2\,m^2_q}{x^2(1-x)^2}\frac{\partial}{\partial M^2_\eta} X_2 - \frac{1}{x(1-x)}\frac{\partial}{\partial M^2_\eta} X_1\right] 
\end{align}

\nt where the parameter dependence has been omitted for brevity, and with the $X$ functions coming from~\cite{Abe:2015rja}
\begin{align}
&\int \frac{\text{d}^{4}k}{(2\pi)^4} \frac{1}{[k^2 - \frac{m_q^2}{x(1-x)}]^n\,[(p_\chi+k)^2 - M^2_X]\,[k^2 - M^2_\eta]}\nn \\[5mm] 
&= \frac{i}{(4\pi)^2}\,X_n\left(p_\chi^2,\,M^2_X,\,M^2_\eta,\,\tfrac{m_q^2}{x(1-x)}\right).
\end{align}


\end{document}